\documentclass[a4paper]{article}

\usepackage[english]{babel}
\usepackage[utf8]{inputenc}
\usepackage[T1]{fontenc}

\usepackage{listings}

\usepackage{lmodern}
\usepackage[autostyle]{csquotes}

\usepackage[a4paper,top=3cm,bottom=2cm,left=3cm,right=3cm,marginparwidth=1.75cm]{geometry}

\usepackage{amsmath}
\usepackage{graphicx}
\usepackage[colorinlistoftodos]{todonotes}
\usepackage{authblk}
\usepackage{rotating}
\usepackage{csvsimple}
\usepackage{float}

\usepackage[colorlinks=true, allcolors=blue]{hyperref}
\usepackage{rotating}

\title{\huge Stool Studies Don't Pass the Sniff Test: \\A Systematic Review of Human Gut Microbiome Research Suggests Widespread Misuse of \\Machine Learning}
\author[1*]{Thomas P. Quinn}

\affil[1]{\footnotesize Applied Artificial Intelligence Institute (A2I2), Deakin University, Geelong, Australia
* \textit{contacttomquinn@gmail.com}
}
\date{}

\Affilfont{\fontsize{4}{4}}

\begin{document}

\maketitle

\begin{abstract}

In the machine learning culture, an independent test set is required for proper model verification. Failures in model verification, including test set omission and test set leakage, make it impossible to know whether or not a trained model is fit for purpose. In this article, we present a systematic review and quantitative analysis of human gut microbiome classification studies, conducted to measure the frequency and impact of test set omission and test set leakage on area under the receiver operating curve (AUC) reporting. Among 102 articles included for analysis, we find that only 12\% of studies report a \textit{bona fide} test set AUC, meaning that the published AUCs for 88\% of studies cannot be trusted at face value. Our findings cast serious doubt on the general validity of research claiming that the gut microbiome has high diagnostic or prognostic potential in human disease.

\end{abstract}

\section{Introduction}

Research suggests that the microbiome, especially the gut microbiome, could serve as a clinical biomarker for disease diagnosis, prognosis, and treatment. These studies use machine learning to fit a model that can predict a clinical outcome as a function of the gut microbe signature. The accuracy of these models are demonstrated through the use of performance metrics like the area under the receiver operating curve (AUC), but can these metrics be trusted?

In the machine learning culture, model verification--also called model validation, as distinct from clinical validation \cite{kim_design_2019}--is used to assess whether a model is fit for purpose. To perform model verification, a model is trained on one set of data, called the \textit{training data}, then tested on completely new unseen data, called the \textit{test data}. If a model performs well for the completely new unseen data, then it can be considered reliable for the tested application \cite{breiman_statistical_2001}.

Model verification is an essential part of a machine learning workflow. When model verification is not performed correctly, or is not performed at all, it becomes impossible to know whether a learned model is accurate \cite{quinn_test_2021}. Two critical errors are
\begin{itemize}
    \item \textbf{Test set omission} When a test set is not used, there are no unseen data available for model verification, meaning that model verification cannot be performed.
    \item \textbf{Test set leakage} When a test set is used, but information from the test set directly or indirectly influences model training, then the test set no longer represents unseen data, meaning that model verification cannot be performed correctly. One example of leakage is selecting features from the combined training and test data, a common machine learning pitfall among genomic studies \cite{teschendorff_avoiding_2019}.
\end{itemize}
Both errors constitute a misuse of machine learning because they bias performance upwards, making a model appear more accurate than it is. They can even make a model appear highly accurate when it is, in fact, not accurate at all. For this reason, one should not take performance metrics at face value; it is necessary to assess the quality of the training regime used to produce them.

This article presents a systematic review and quantitative analysis of the human gut microbiome literature that seeks to measure the frequency and impact of \textbf{test set omission} and \textbf{test set leakage} on AUC reporting. We find evidence that the absence or misuse of test sets is widespread, with only 12\% of studies performing model verification correctly. Moreover, the 88\% of studies that fail to perform model verification report significantly higher AUCs (95\% CI: 2.3 to 11.1 point difference). Our findings cast serious doubt on the general validity of research claiming that the gut microbiome has high diagnostic or prognostic potential in human disease.

\section{Methods}

\subsection{Study retrieval and screening}

On June 2, 2021 we searched PubMed using the term ``((GUT) OR (STOOL) OR (FECAL) OR (FAECAL)) AND (MICROB*) AND (AUC)''. We had 4 inclusion criteria: (1) the study involved human subjects, (2) the abundances of their gut microbes were measured by a high-throughput assay (e.g., microarray, 16s, or shotgun sequencing), (3) the microbe abundances were used as features to train a classifier, and (4) the AUC of the classifier was reported in the Abstract. Criteria 4 was included because Abstract-reported AUC is the outcome for our quantitative analysis.

We had 7 exclusion criteria: (1) microbe abundances were not measured using a high-throughput assay (e.g., qPCR); (2) microbe abundances were not used as features to train a classifier; (3) fungal or viral abundances were measured without also measuring bacterial microbe abundances; (4) metabolite abundances were measured without also measuring bacterial microbe abundances; (5) reviews or systematic reviews reporting meta-analyses; (6) methodological studies that propose a new tool or that benchmark several competing tools; or (7) article not available in English (2 articles) or not available from our University library (1 article).

\begin{figure}[H]
\centering
\scalebox{1}{
\includegraphics[width=(.5\textwidth)]{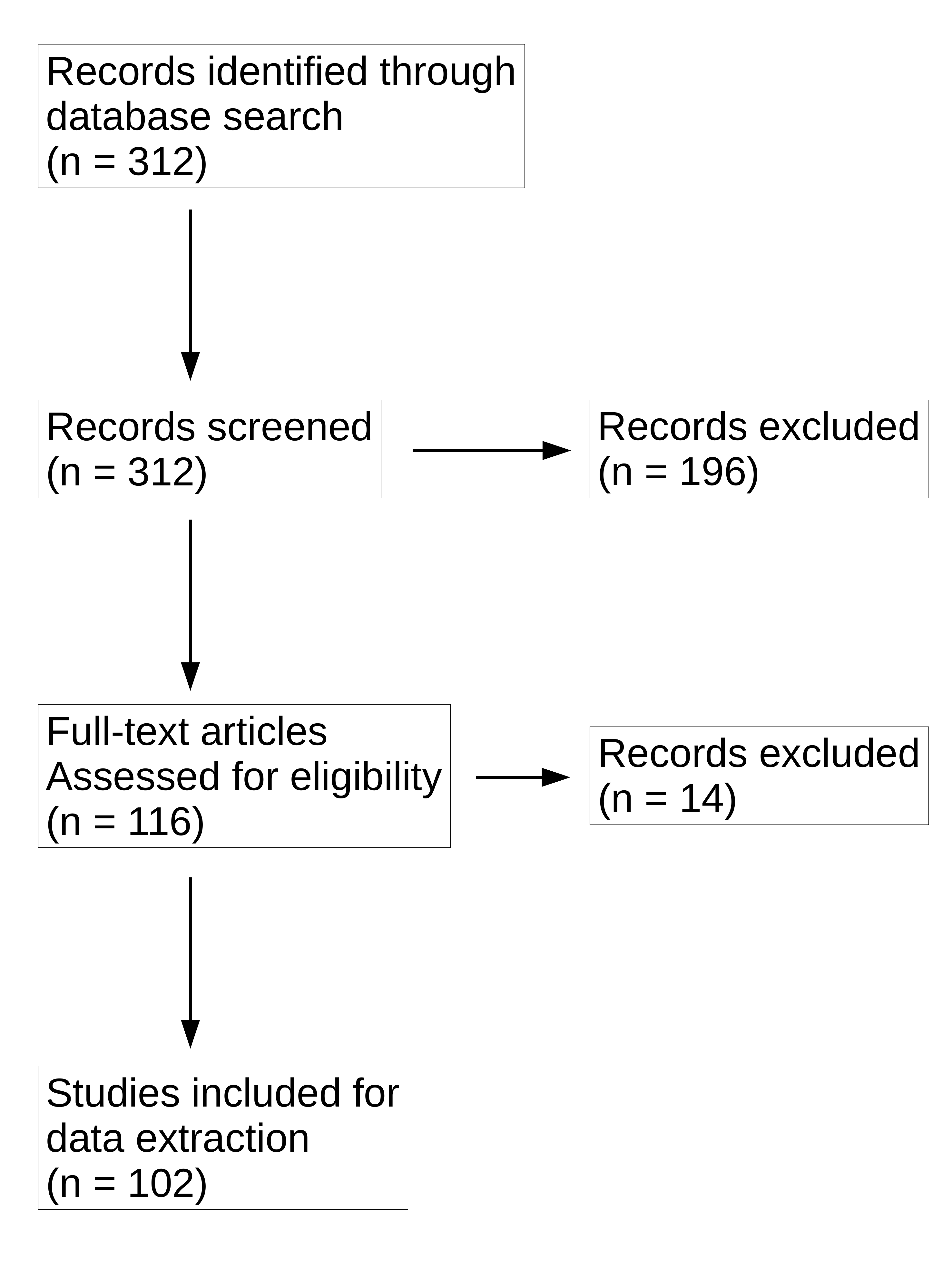}}
\caption{A workflow of study retrieval, screening, and exclusion. Overall, we included 102 studies for data extraction and quality assessment.}
\label{fig:datypes}
\end{figure}

\subsection{Data extraction}

We extracted 5 pieces of data from Abstracts.
\begin{itemize}
    \item The study DOI.
    \item The first affiliated location of the senior author.
    \item Whether the researchers conclude explicitly in the Abstract that their classifier can be used, or has promise to be used, as a biomarker-based prediction tool (Yes or No).
    \item Whether the researchers conclude explicitly in the Abstract that their classifier cannot be used as a biomarker-based prediction tool (Yes or No).
    \item The Abstract-reported AUC as measured on the \textit{most independent data set} (i.e., rank order: external test set $>$ internal test set $>$ validation set $>$ training set). In the case of multiple classification tasks, we record the highest independent AUC. For example, if the authors published two training set AUCs of 91\% and 95\%, and two validation set AUCs of 84\% and 89\%, we would record 89\%.
\end{itemize}
We extract an additional 5 pieces of data from the Methods.
\begin{itemize}
    \item Whether the study meets each of the 4 inclusion criteria.
    \item The positive and negative classes for the classification task. In the case of multiple tasks, we record the task that corresponds to the Abstract-reported AUC.
    \item The total sample size, as well as the sample size for each group. This sample size includes all patients from training, validation, and test sets. In the case of multiple classification tasks, we record the sample sizes that correspond to the Abstract-reported AUC.
    \item The assay used to measure gut microbe abundances.
    \item The classification algorithm used.
\end{itemize}

\subsection{Quality assessment}

We assess the quality of each article based on 5 criteria. All criteria recorded as ``1'' (Done and reported clearly) or ``0'' (Not done or not reported clearly).
\begin{itemize}
    \item \textbf{VAL\_USE} Records whether a validation set is used, including cross-validation. This assessment is based on our judgement, not the terminology used by the authors.
    \item \textbf{TEST\_USE} Records whether an \textit{internal} test set is used. Unlike a validation set, a test set is set aside before model training. By internal, we mean that the test set comes from the same patient cohort used for model training. This assessment is based on our judgement, not the terminology used by the authors. Set to ``1'' if nested cross-validation, or cross-validation without feature selection, is performed. 
    \item \textbf{EXT\_USE} Records whether an \textit{external} test set is used. By external we mean that the test set comes from a second patient cohort. External test sets must be described as being collected separately from the training, validation, and internal test sets.
    \item \textbf{REPORT\_TEST} Records whether AUC reported in the Abstract comes from a test set (internal or external). This is not an assessment of whether the test set has leaked.
    \item \textbf{NO\_LEAK} Records whether feature selection is performed on the training data only. In other words, this records whether feature selection happens independently from the test set, without test set leakage. Set to ``1'' if no feature selection is performed.
\end{itemize}
In addition to these 5 quality assessments, we record a final assessment of the article by answering the statement ``I am confident that the AUC reported in the article is calculated on an independent test set that has not leaked into the training of the model.'' Answers are recorded as ``1'' (I agree) or ``0'' (I do not agree).

Quality assessments and data extractions are performed based on a reading of the Methods or equivalent section. Supplemental Methods were not read unless the entire Methods were provided as a Supplemental file. If we could not make the final assessment based on the Methods alone, we would refer to the Results section to see if it contained additional methodological details.

\subsection{Quantitative analysis}

Associations between AUC and sample size were evaluated by Pearson correlation. Associations between AUC and quality assessments were evaluated by Student's t-test. All prior $\alpha=0.05$.

\section{Results}

\subsection{Study characteristics}

We included 102 studies for data extraction. The studies overwhelmingly used 16s sequencing (n=85), followed by shotgun sequencing (n=14), phylogenetic array (n=2), and multiple platforms (n=1). The most represented geographic regions were China and Taiwan (n=63), followed by Europe (n=20), North America (n=13), Near East (n=3), Far East (n=2), and Oceania (n=1).

The medical conditions under study varied greatly. After extracting all data, we assigned each study a category based on the organ system involved in the condition, but giving colorectal cancer (CRC) and inflammatory bowel disease (IBD) their own categories (see Table~\ref{tab:cats}). The most represented categories were CRC (n=15), IBD (n=12), and other enteric conditions (n=15). Autoimmune or inflammatory disease (n=10), neurology (n=9), and psychiatry (n=8) were also popular. Of note, 3 studies did not research a specific medical condition, but rather sampled the general population or a cohort of healthy subjects.

\begin{table}[h]
\centering
\begin{tabular}{|l|l|}
\hline
\textbf{Condition Under Study} & \textbf{Count}\\
\hline
enteric & 15\\
\hline
colorectal cancer & 15\\
\hline
inflammatory bowel disease & 12\\
\hline
autoimmune or inflammatory & 10\\
\hline
neurologic & 9\\
\hline
psychiatric & 8\\
\hline
infectious & 7\\
\hline
diet, nutrition, or obesity & 7\\
\hline
renal or diabetes & 7\\
\hline
other & 5\\
\hline
cardiac & 4\\
\hline
none & 3\\
\hline
\end{tabular}
\caption{The types of studies included for data extraction and quality assessment, classified by the organ system involved in the condition, but giving colorectal cancer (CRC) and inflammatory bowel disease (IBD) their own categories.}
\label{tab:cats}
\end{table}

\subsection{Test sets overwhelmingly absent or leaky}

Test sets are overwhelmingly absent. Only 21 out of 102 studies used an internal test set that was set aside from the training data.
Of these, only 14 did not leak the test data during feature selection. Of these, only 11 reported the AUC for the test set (the others ostensibly reporting AUC for all samples, training and test sets combined). Overall, only 11 of the 102 studies reported a \textit{bonda fide} test set AUC. An additional study did not use a test set \textit{per se}, but replicated a previously published biomarker classifier. These 12 are the only 12 studies that passed our final assessment: ``I am confident that the AUC reported in the article is calculated on an independent test set that has not leaked into the training of the model.''

\begin{figure}[H]
\centering
\scalebox{1}{
\includegraphics[width=(.6\textwidth)]{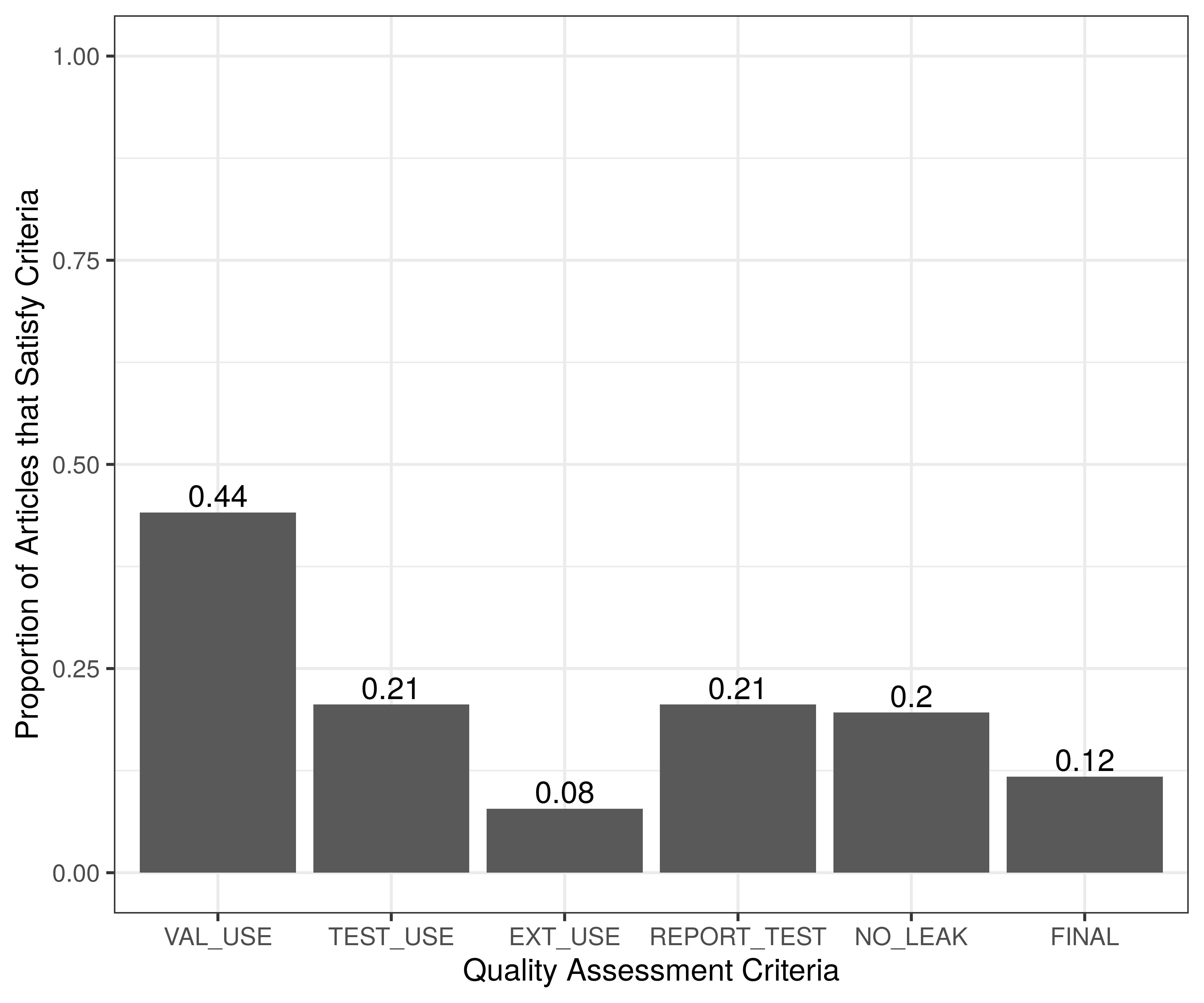}}
\caption{The proportion of studies meeting each of the 6 quality assessments, including our final assessment: ``I am confident that the AUC reported in the article is calculated on an independent test set that has not leaked into the training of the model.''}
\label{fig:props}
\end{figure}

Figure~\ref{fig:props} shows the proportion of articles satisfying each of the quality assessment criteria, along with the final assessment. Impressively, more than half of articles surveyed did not include \textit{any} validation (e.g., cross-validation), even if leaky. Meanwhile, only 8\% of studies used an \textit{external} test set collected separately from the training data, meaning that most studies lacked the data needed to demonstrate model generalizability, an important step towards establishing clinical validity.

\subsection{Studies without test sets report higher AUC}

\begin{figure}[H]
\centering
\scalebox{1}{
\includegraphics[width=(.6\textwidth)]{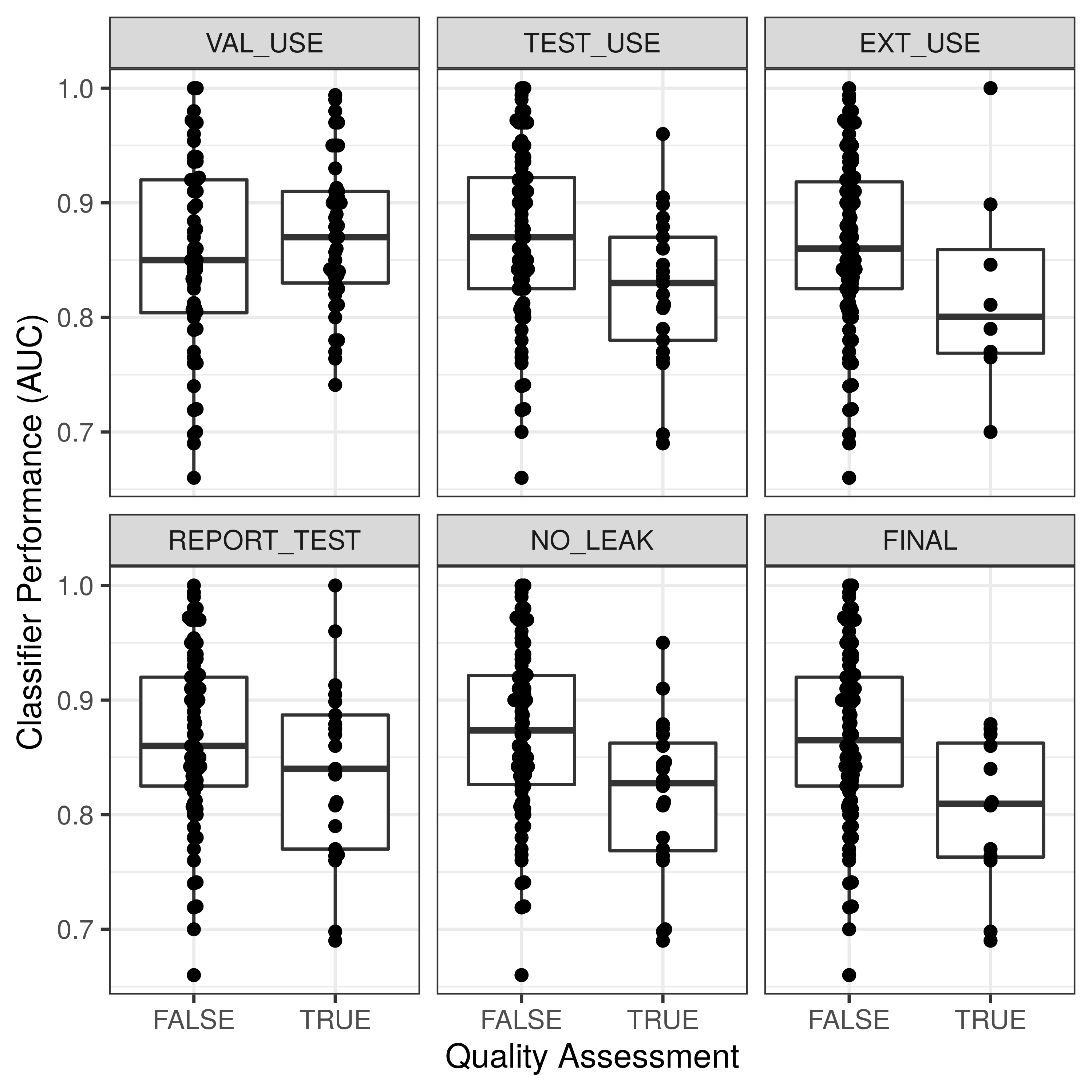}}
\caption{The distribution of Abstract-reported AUCs for studies that did and did not meet each quality assessment criteria. Studies that fail to meet certain quality assessment criteria also tended to report higher AUCs in the Abstract. Mean differences for criteria \textbf{TEST\_USE}, \textbf{NO\_LEAK}, and \textbf{FINAL} all $p<0.05$ by t-test.}
\label{fig:aucqc}
\end{figure}

Studies that did not use test sets reported higher AUCs than studies that did (95\% CI: 1.3 to 8s.1 point difference). Meanwhile, studies that leaked the test data reported higher AUCs than studies that did not (95\% CI: 2.1 to 9.3 point difference). These differences were even greater when comparing studies that did and did not pass our final assessment (95\% CI: 2.3 to 11.1 point difference).
We observed no significant differences based on whether studies used validation sets, used external test sets, or claimed to report test set AUC. Figure~\ref{fig:aucqc} shows the relevant box plots.

\subsection{Smaller studies report higher AUC}

\begin{figure}[H]
\centering
\scalebox{1}{
\includegraphics[width=(.6\textwidth)]{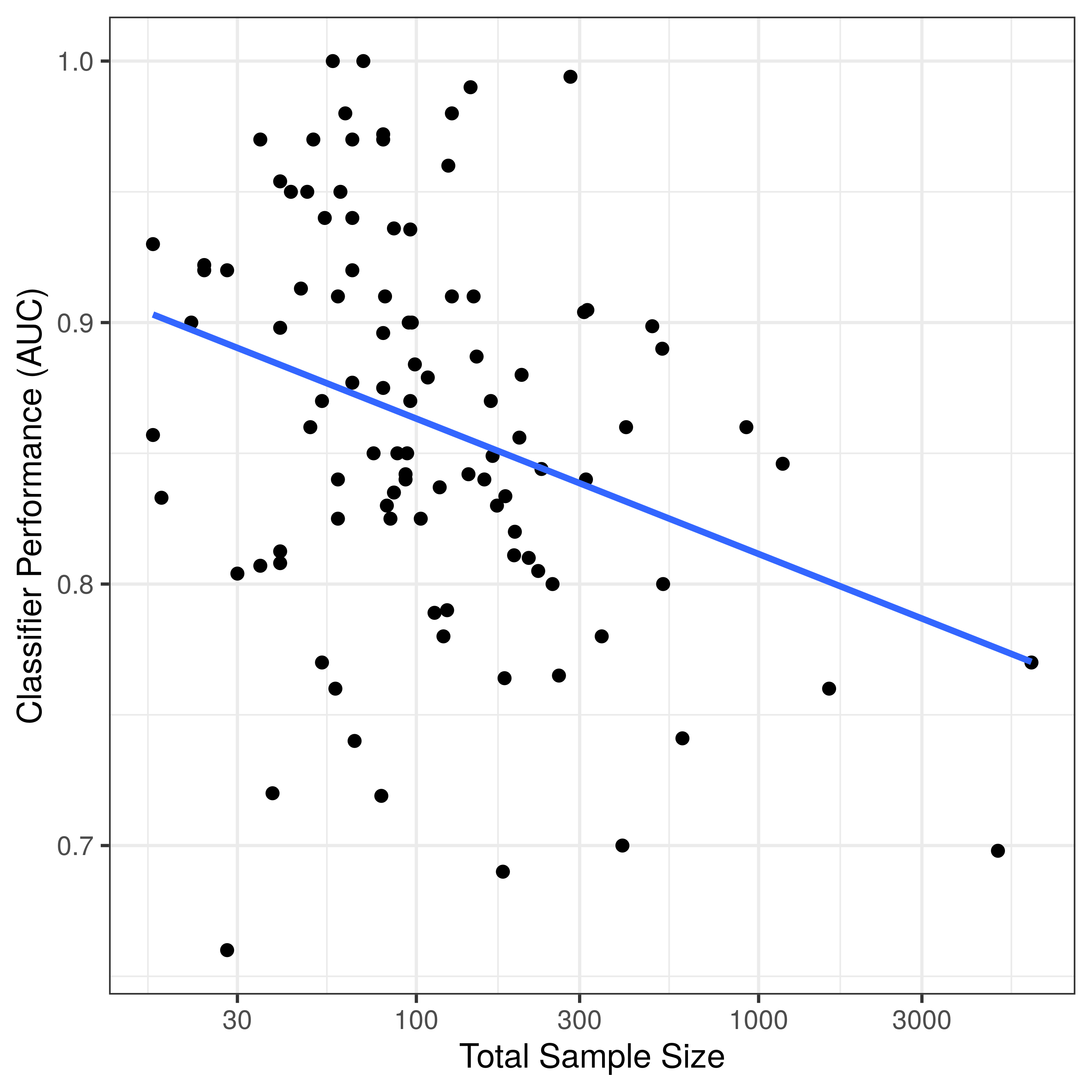}}
\caption{A scatter plot of the Abstract-reported AUC (y-axis) as a function of sample size (x-axis). Studies with larger sample sizes tended to report lower AUCs ($\rho=-0.31$; $p=0.0013$).}
\label{fig:cor}
\end{figure}

Reported AUCs range from 0.56 to 1, with a median of 0.8585. 34\% of studies reported an AUC at or above 0.9, while 15.5\% of studies reported an AUC at or above 0.95. AUCs were not distributed randomly with respect to sample size. Studies with larger sample sizes tended to report lower AUCs ($\rho=-0.31$; $p=0.0013$). This is noteworthy because, all things being equal, machine learning fundamentals would suggest that having more data would result in a \textit{higher} AUC because it becomes easier to find a robust and generalizable signal among the noise. The negative correlation between AUC and sample size contradicts this expectation. Figure~\ref{fig:cor} shows the association between classifier performance and log-scale sample size.

\section{Discussion}

AUCs range from 0 to 1, where a 1 indicates that a model always makes the correct prediction. High AUCs, like small p-values, may be used to convince readers that the experimental results are ``significant''. However, also like p-values, the interpretation of an AUC depends on the context in which it was measured. Two key errors, \textbf{test set omission} and \textbf{test set leakage}, can make published AUCs unreliable at best, and misleading at worst. Our Results suggest that the AUC from 88\% of human gut microbiome classification studies cannot be trusted at face value. To contextualize this finding, let us imagine we select a study at random and then flip 3 coins. It is more likely that all coins will land heads than that the reported AUC can be trusted. 

Moreover, we find a significant difference in AUC between studies that did use and did not use test sets correctly. We interpret this as evidence that the widespread misuse of machine learning has resulted in overly optimistic AUCs. Although we measure the magnitude of difference in AUC as only 2.3-11.1 points, we speculate that the true effect size of \textbf{test set omission} and \textbf{test set leakage} is much higher. This is because the \textit{studies correctly using test sets} are not counter-factuals of the \textit{studies incorrectly using test sets}. Rather, the AUCs reported by the \textit{studies correctly using test sets} may represent some of the highest possible AUCs because of a publication bias that puts upward pressure on published AUCs (i.e., analogous to the publication bias that puts downward pressure on published p-values). Indeed, only 1/102 studies reported the null finding that the classifier they trained performed no better than chance. Yet, it seems unlikely that 99\% of all experimental gut microbiome classifiers would work out on the first try. 

Our Results contribute to a growing literature that exposes systemic problems with the quality of microbiome research and clinical machine learning. Among these, Nearing et al. showed, through benchmark exercises, that different differential abundance software tools produce disturbingly different results, with many failing to control false discovery rate (FDR) \cite{nearing_microbiome_2021}, replicating findings from \cite{hawinkel_broken_2017}. Meanwhile, Walter et al. found that 95\% of studies investigating the rodent microbiome inappropriately generalized findings across species to make unfounded causal inferences about human disease \cite{walter_establishing_2020}. Outside of microbiome research, a recent systematic review of medical AI applications found widespread exaggeration about the clinical utility of machine learning models, alongside routine study bias, opaque reporting, and poor data availability \cite{nagendran_artificial_2020}. These findings all raise concerns that new technologies, while enabling new experiments, may bring them new threats to reproducibility, replicability, robustness, and generalizability \cite{schloss_identifying_2018}.

Readers who would like to learn more about machine learning best practices can refer to the helpful introductory works by Davide Chicco \cite{chicco_ten_2017} and Andrew Teschendorff \cite{teschendorff_avoiding_2019}, as well as introductory coding textbooks like those published by O'Reilly. When in doubt, we encourage our readers to reach out to local computational biology experts for questions and support.

\section{Limitations}

Finally, we ask our readers to interpret our Results in light of two limitations.
First, only one reviewer engaged in Abstract screening, data extraction, and quality assessment. Despite earnest attempts to define the data extraction and quality assessment criteria clearly, there nonetheless remains some subjectivity to the process.
Second, we focus on gut microbiome classification studies that report AUC in the Abstract, which is a subset of all gut microbiome classification studies. Although we use this inclusion criteria intentionally in order to define an outcome for our quantitative analysis, we acknowledge that it may introduce a selection bias. Notably, classification studies \textit{not} reporting AUC in the Abstract may be more likely to use machine learning correctly, for example because machine learning experts often prefer the F1-score over the AUC. Given these limitations, as well as the gravity of the Results themselves, we encourage others to consider replicating this kind of analysis in similar gut microbiome and clinical biomarker studies.

\section{Conclusions}

Plug-and-play analytical software has made it easier than ever to apply machine learning to biological data. Unfortunately, this has also made it easy for amateurs to misuse machine learning. In this study, we show that common errors like \textbf{test set omission} and \textbf{test set leakage} are widespread, occurring in 88\% of surveyed human gut microbiome classification studies. Our findings cast serious doubt on the general validity of research claiming that the gut microbiome has high diagnostic or prognostic potential in human disease.

\section{Availability of data and code}

Data and code available from \url{https://doi.org/10.5281/zenodo.5069117}.

\bibliographystyle{unsrt}
\bibliography{references}

\begin{thebibliography}{10}

\bibitem{kim_design_2019}
Dong~Wook Kim, Hye~Young Jang, Kyung~Won Kim, Youngbin Shin, and Seong~Ho Park.
\newblock Design {Characteristics} of {Studies} {Reporting} the {Performance}
  of {Artificial} {Intelligence} {Algorithms} for {Diagnostic} {Analysis} of
  {Medical} {Images}: {Results} from {Recently} {Published} {Papers}.
\newblock {\em Korean Journal of Radiology}, 20(3):405--410, March 2019.

\bibitem{breiman_statistical_2001}
Leo Breiman.
\newblock Statistical {Modeling}: {The} {Two} {Cultures} (with comments and a
  rejoinder by the author).
\newblock {\em Statistical Science}, 16(3):199--231, August 2001.

\bibitem{quinn_test_2021}
Thomas~P. Quinn, Vuong Le, and Adam P.~A. Cardilini.
\newblock Test set verification is an essential step in model building.
\newblock {\em Methods in Ecology and Evolution}, 12(1):127--129, 2021.

\bibitem{teschendorff_avoiding_2019}
Andrew~E. Teschendorff.
\newblock Avoiding common pitfalls in machine learning omic data science.
\newblock {\em Nature Materials}, 18(5):422--427, May 2019.

\bibitem{nearing_microbiome_2021}
Jacob~T. Nearing, Gavin~M. Douglas, Molly Hayes, Jocelyn MacDonald, Dhwani
  Desai, Nicole Allward, Casey M.~A. Jones, Robyn Wright, Akhilesh Dhanani,
  André~M. Comeau, and Morgan G.~I. Langille.
\newblock Microbiome differential abundance methods produce disturbingly
  different results across 38 datasets.
\newblock {\em bioRxiv}, page 2021.05.10.443486, May 2021.

\bibitem{hawinkel_broken_2017}
Stijn Hawinkel, Federico Mattiello, Luc Bijnens, and Olivier Thas.
\newblock A broken promise: microbiome differential abundance methods do not
  control the false discovery rate.
\newblock {\em Briefings in Bioinformatics}, August 2017.

\bibitem{walter_establishing_2020}
Jens Walter, Anissa~M. Armet, B.~Brett Finlay, and Fergus Shanahan.
\newblock Establishing or {Exaggerating} {Causality} for the {Gut}
  {Microbiome}: {Lessons} from {Human} {Microbiota}-{Associated} {Rodents}.
\newblock {\em Cell}, 180(2):221--232, January 2020.

\bibitem{nagendran_artificial_2020}
Myura Nagendran, Yang Chen, Christopher~A. Lovejoy, Anthony~C. Gordon, Matthieu
  Komorowski, Hugh Harvey, Eric~J. Topol, John P.~A. Ioannidis, Gary~S.
  Collins, and Mahiben Maruthappu.
\newblock Artificial intelligence versus clinicians: systematic review of
  design, reporting standards, and claims of deep learning studies.
\newblock {\em BMJ}, 368, March 2020.

\bibitem{schloss_identifying_2018}
Patrick~D. Schloss.
\newblock Identifying and {Overcoming} {Threats} to {Reproducibility},
  {Replicability}, {Robustness}, and {Generalizability} in {Microbiome}
  {Research}.
\newblock {\em mBio}, 9(3):e00525--18, 2018.

\bibitem{chicco_ten_2017}
Davide Chicco.
\newblock Ten quick tips for machine learning in computational biology.
\newblock {\em BioData Mining}, 10(1):35, December 2017.

\end{thebibliography}

\end{document}